\documentclass[12pt]{article}
\usepackage{times,epsfig,cite,amssymb}

\normalbaselines
\title{Coherent states for quadratic Hamiltonians}

\author{Alonso Contreras-Astorga, David J. Fern\'andez C. and Mercedes Vel\'azquez
\\ [12pt] {\sl Departamento de F\'{\i}sica, Cinvestav, A.P. 14-740,
07000 M\'exico D.F., Mexico}}

\date{}

\begin{document}
\maketitle

\begin{abstract}
The coherent states for a set of quadratic Hamiltonians in the trap regime are
constructed. A matrix technique which allows to identify directly the creation and annihilation
operators will be presented. Then, the coherent states as simultaneous eigenstates
of the annihilation operators will be derived, and they are going to be compared
with those attained through the displacement operator method. The corresponding wave function
will be found, and a general procedure for obtaining several mean values involving
the canonical operators in these states will be described. The results will be illustrated
through the asymmetric Penning trap.
\end{abstract}

{PACS: 03.65.Ge, 03.65.Sq, 37.10.Ty, 37.30.+i}

\section{Introduction}

As it was shown by Schr\"odinger in 1926 for the harmonic oscillator,
the quasi-classical states are important for the description of physical
systems in the classical limit (see e.g. \cite{Cohen1}).
The catchy term {\it coherent states} (CS)
was used for the first time by Glauber long after,
when studying electromagnetic correlation functions \cite{GlauberFeb63,GlauberSep63}.
With this application it was realized that the CS are
useful as well in the intrinsically quantum domain. Indeed, the CS approach is nowadays
widely employed for dealing with quantum physical systems. According to Glauber
there are three equivalent ways to construct the CS for the harmonic oscillator.
The first one is to define them as eigenstates of the annihilation operator. The second
one is to build the CS through the action of a displacement operator onto the ground
state. The third way is to consider them as quantum states having a minimum Heisenberg
uncertainty relationship. These three properties can be used as definitions to build
the CS for systems different from the harmonic oscillator. However,
it is noteworthy that each of them leads to sets of CS which do not coincide in general
\cite{Bargman,
generalizacionECLiecompactos,
generalizacionECgruposLie,
generalizacionECRelacionesIncertidumbre,
PerelomovBook,
libroCompliacion,zfg90,
NuevoDelProfe,qu01,do02}.
In fact, even for the harmonic oscillator the third definition does not produce just
the standard CS, since it includes as well the so-called squeezed states \cite{tesisLic,fr06}.

In spite of its long term life, from time to time there are some advances which
maintain the subject alive. This is the case, e.g., of the recently discovered coherent
states for a charged particle inside an ideal Penning trap \cite{fv09}. Since the
corresponding Hamiltonian is quadratic in the position and momentum operators, one would
expect that the CS appear as a generalized displacement operator acting onto the
corresponding ground state. However, it is worth to notice that the Penning trap
Hamiltonian does not have any ground state at all, since it is not a positively defined operator.
Despite, it was possible to implement in a simple way the corresponding CS construction.
Thus, we need to take into account this Hamiltonian property when studying the coherent states
of general systems.

In this article we are going to address the CS construction for systems characterized by a
certain set of quadratic Hamiltonians. The CS will be built up as simultaneous
eigenstates of the corresponding annihilation operators, and also by applying
a generalized displacement operator onto an appropriate extremal state.
We will see that in the case of a positively defined Hamiltonian this extremal state will
coincide with the ground state. In order to perform the CS construction,
we need to find first the annihilation and creation operators. This task
will be done by using a matrix technique, which generalizes the one employed in \cite{fv09}
(see also \cite{dm89,ve07,mr10a,mr10b}).
In this way, we will simply and systematically identify the characteristic
algebra of the involved Hamiltonians. Our procedure represents a generalization to dimensions
greater than one of the standard technique to deal with the harmonic oscillator, which is
closely related to the well known factorization method (see, e.g., \cite{mr04,ff05}).

The paper is organized as follows. In section 2 we will introduce a detailed recipe for
systematically obtaining the annihilation and creation operators for quadratic
Hamiltonians in the trap regime. The coherent states derivation shall be elaborated in
section 3, while in section 4 we will address the completeness of this set of CS,
we shall obtain the mean values of several important physical quantities and the time evolution of these states. We are going to
apply our general results to an asymmetric Penning trap in section 5, and our conclusions
will be presented at section 6.

\section{Ladders operators for quadratic Hamiltonians}\label{S:evol temp}

Along this work we are going to consider a general set of $n$-dimensional
quadratic Hamiltonians of the form
\begin{equation}\label{hamiltoniano gral}
H=\frac{1}{2}\eta^T \mathbf{B}\eta ,
\end{equation}
where $\mathbf{B}$ is a $2n\times2n$ real constant symmetric matrix,
$\eta=\left(\begin{array}{c} \vec{X}, \vec{P} \end{array} \right)^T$,
and $\vec{X}, \vec{P}$ are the $n$-dimensional coordinate and
momentum operators in the Schr\"odinger picture satisfying the
canonical commutation relationships
$[X_i,P_j] = i \delta_{ij}$ (notice that a system of units such that $\hbar=1$
will be used throughout this paper). The time evolution of the operator vector
$\eta(t) = U^\dagger(t) \eta U(t)$ in the Heisenberg picture is governed by
\begin{equation}\label{evolu temp}
\frac{d\eta(t)}{dt}= U^\dagger(t)[iH, \eta]U(t) =
U^\dagger(t)\mathbf{\Lambda} \eta U(t) =
    \mathbf{\Lambda}\eta(t),
\end{equation}
$U(t)$ being the evolution operator of the system such that $U(0) = {\bf 1}$, and
$\mathbf{\Lambda}=\mathbf{JB}$, where $\mathbf{J}$ is the well known $2n\times2n$
matrix
\begin{equation}
    \mathbf{J}=\left(\begin{array}{cc} 0 & \mathbf{1}_n\\
                                    -\mathbf{1}_n& 0
                                    \end{array}\right),
\end{equation}
satisfying
\begin{equation}\label{propiedades J}
    \mathbf{J}^T=-\mathbf{J},\qquad \quad
    \mathbf{J}^2=-\mathbf{1}_{2n},\qquad \quad
    {\rm det}(\mathbf{J})=1,
\end{equation}
in which $\mathbf{1}_m$ represents the $m\times m$ identity matrix.
The solution of Eq.(\ref{evolu temp}) is given by
\begin{equation}
    \eta(t)=e^{\mathbf{\Lambda}t}\eta(0)= e^{\mathbf{\Lambda}t}\eta.
\end{equation}

In order to identify the annihilation and creation operators of $H$, we
need to find the right and left eigenvectors of $\mathbf{\Lambda}$.
Since in general $\mathbf{\Lambda}$ is non-hermitian, its right and left
eigenvectors are not necessarily adjoint to each other.

Let us consider in the first place the $2n$-th order characteristic polynomial of
$\mathbf{\Lambda}$,
 $P(\lambda)={\rm det}(\mathbf{\Lambda} -
\lambda) = {\rm det}(\mathbf{J}\mathbf{B}-\lambda)$. Using Eqs.(\ref{propiedades J})
and the fact that $\mathbf{B}$ is a symmetric matrix, we obtain
\begin{equation}
P(\lambda)  =  {\rm det}(\mathbf{J}\mathbf{B}-\lambda)  =  {\rm det}[(\mathbf{J}\mathbf{B}-\lambda)^T]
 =  {\rm det}(\mathbf{J}\mathbf{B} + \lambda)  =  P(-\lambda).
\end{equation}
This means that if $\lambda$ is an eigenvalue of $\mathbf{\Lambda}$,
then $-\lambda$ also will be. Throughout this work we are going to denote
the eigenvalues of $\mathbf{\Lambda}$ as $\lambda_k$ and $-\lambda_k$, taking
$\lambda_k$ in the way
\begin{equation}\label{convencion}
{\rm{Re}}(\lambda_k)  >  0
\quad \textnormal{or} \quad
{\rm{Im}}(\lambda_k)  >  0 \  \textnormal{if} \
{\rm{Re}}(\lambda_k) =  0, \quad k  =  1,\dots,n.
\end{equation}
Let us label as $u_k^\pm$ and $f_k^\pm$ the right and left
eigenvectors associated to the eigenvalues $\pm \lambda_k$
respectively, i.e.,
\begin{equation}\label{evefl}
    \mathbf{\Lambda} u_k ^\pm = \pm \lambda_k u_k ^\pm ,
    \qquad
    f_k ^\pm \mathbf{\Lambda} = \pm \lambda_k  f_k ^\pm.
\end{equation}
Notice that both $u_k^\pm$ and $f_k^\pm$ can be determined from
Eq.(\ref{evefl}) up to arbitrary factors. Part of this arbitrariness will
be eliminated by imposing two requirements. The first one is that the
right and left eigenvectors be dual to each other, namely,
\begin{equation}\label{normalizacion a deltas}
f_j^r u_k ^{r'} = \delta_{jk} \delta_{rr'},
\end{equation}
with $j,k=1,\dots,n$ and $r,r'= +, -$. The second condition, which is needed
in order to recover the standard annihilation and creation
operators for the one-dimensional harmonic oscillator, is to ask that the
left eigenvectors $f_k^\pm$ involved in the commutators
\begin{equation}\label{segundacondicion}
[f_k^- \eta , f_k^+ \eta] = \gamma_k, \quad \gamma_k \in {\mathbb C},
\end{equation}
are such that
\begin{equation}\label{modulogamma}
\vert \gamma_k \vert = 1 .
\end{equation}

In this paper we are going to discuss just the case in which there are no
degeneracies in the eigenvalues $\pm \lambda_k$ so that the identity matrix
$\mathbf{1}_{2n}$ can be expanded as (see \cite{Faisal})
\begin{equation}
    \mathbf{1}_{2n} = \sum_{k=1}^{n} \big( u_k^+\otimes f_k^+ +
                            u_k^-\otimes f_k^-
                            \big),
\end{equation}
with $\otimes$ representing tensor product. Then we get
\begin{eqnarray}\nonumber
   && \hskip-1.0cm  \eta(t) = 
        e^{\mathbf{\Lambda}t} \left[ \sum_{k=1}^{n} \big( u_k^+ \otimes f_k^+ +
                                         u_k^- \otimes f_k^-
                                    \big) \right] \eta
       \! = \! \! \sum_{k=1}^n \big[e^{\lambda_kt} u_k^+ \otimes f_k^+ \eta +
                            e^{-\lambda_kt} u_k^- \otimes f_k^- \eta
                    \big]
    \\
    \label{eta(t)}
        && \hskip-0.3cm
        = \sum_{k=1}^n \big( e^{\lambda_kt} u_k^+  L_k^+
                        +e^{-\lambda_kt} u_k^- L_k^-
                        \big),
\end{eqnarray}
where $L_k^\pm \equiv f_k^\pm \eta$.

It is worth to point out that, in the classical case, $L_k^\pm$ represent
$c$-numbers which are related with the initial conditions. The
time dependence of $\eta(t)$ is determined essentially by the $\lambda_k$-values,
which are complex in general. If ${\rm{Re}}(\lambda_k)\neq 0$ it can be seen that
one of the two involved exponentials of the $k$-th term in the previous relation
diverges as $t$ increases and, thus, the classical motion will be in general unbounded
(see \cite{camporotante,Chico,tesisSara,cm06}). The only way in which this does not happen is
that all the eigenvalues be purely imaginary so that the corresponding exponentials will
just induce oscillations in time, and therefore in this case the classical evolution of the vector
$\eta(t)$ will remain always bounded.

On the other hand, in the quantum regime the $L_k^\pm$ are linear operators
in the canonical variables $\vec{X}, \vec{P}$. It is straightforward to show
that their commutators with $H$ reduce to
\begin{equation}\label{conmutador con H}
    [H, L_k ^\pm] = \mp i \lambda_k L_k ^\pm.
\end{equation}
In addition, it turns out that
\begin{eqnarray}\label{conmutadores1}
   & [L_j ^- , L_k ^- ] = [L_j ^+ , L_k ^+] = 0, \quad
     [L_j ^- , L_k ^+] = 0, \quad k\neq j.
\end{eqnarray}
However,
\begin{equation}\label{conmutadores2}
[L_k ^- , L_k ^+ ] = \gamma_k \neq 0, \quad k = 1,\dots, n.
\end{equation}
Eq.(\ref{conmutador con H}) implies that $L_k ^\pm$ behave, at least formally, as 
ladders operators for the eigenvectors of $H$, changing its eigenvalues by $\mp i\lambda_k$. 
However, this statement has to be managed carefully since it could happen that the action of 
$L_k ^\pm$ onto an eigenvector of $H$ produces something which does not belong to the 
domain of $H$, and in this case we would not get new eigenvectors of $H$. In the next section 
we will explore an interesting situation (the trap regime of our systems) for which the
application of $L_k ^\pm$ onto an eigenvector of $H$ produce a new one with different 
eigenvalue.

Let us point out that Eqs.(\ref{conmutador con H}-\ref{conmutadores2})
imply that $H$ can be expressed in a simple way in terms of $\{L_k^\pm,
k=1,\dots,n\}$ \cite{camporotante}. This is a consequence of the following
general theorem:

\medskip

\noindent {\it Theorem}. If $\mathcal{L}$ is an irreducible algebra of operators
generated by $L_i^{\pm}$ which obey $[L_i^+,L_j^+]=[L_i^-,L_j^-]=0$,
$[L_i^-,L_j^+]=\gamma_i\delta_{ij}$ with $|\gamma_i|=1,$ $i,j=1,\dots,n$, then an
operator $H\in \mathcal{L}$ which fulfills relation (\ref{conmutador con H}) can be
written as
\begin{equation}\label{teorema g0}
H=\sum_{k=1}^n \bigg( \frac{-i\lambda_k}{\gamma_k}\bigg) L^+_k
L^-_k + g_0,
\end{equation}
where $g_0\in{\mathbb C}$.

\smallskip

\noindent {\it Demonstration}. Actually, due to Eqs.(\ref{conmutador con H}-\ref{conmutadores2})
it turns out that
$$
g_0 \equiv H - \sum_{k=1}^n \bigg( \frac{-i\lambda_k}{\gamma_k}\bigg) L^+_k L^-_k
$$
commutes with $L_k^\pm$ for all $k$, thus with any
function of them, and so $g_0$ must be a $c$-number.$\square$

\medskip

From now on we are going to discard situations such that ${\rm Re}(\lambda_k)\neq 0$
for some $k=1,\dots,n$, restricting ourselves to cases in which the $\lambda_k$ are
purely imaginary for all $k$, i.e., we stick to the trap regime of our systems.

\subsection{Algebraic structure of $H$ in the trap regime}\label{secc.imaginarios}

Let us suppose that $\lambda_k=i \omega_k$ with $\omega_k > 0,$
$k=1,\dots,n$. Hence, $\lambda_k ^*
 =-\lambda_k$, and since $\mathbf{\Lambda}$ is real, without loosing
generality we can choose
 \begin{equation}
f_k ^- = (f_k ^+)^*, \qquad u_k ^- = (u_k ^+)^*.
 \end{equation}
Since $L_k^\pm$ are linear combinations of the hermitian components of
$\eta$ ($X_1,\dots,X_n,$ $P_1,\dots,P_n$), it turns out that
\begin{equation}
(L_k ^\pm)^\dag = L_k^\mp.
\end{equation}
Note moreover that $\gamma_k^* = \gamma_k$, i.e., $\gamma_k \in {\mathbb R}$, and
the use of Eq.(\ref{modulogamma}) implies that $\gamma_k = \pm 1$. Summarizing
these results, Eq.(\ref{conmutador con H}) becomes in this case
\begin{equation}\label{A+, A-_ operadores de escalera de H}
    [H,L_k^\pm] = \pm \omega_k L_k^\pm,
\end{equation}
i.e., the modifications suffered by the eigenvalues of $H$ through
the action of the ladder operators $L_k^\pm$ are given by the real
quantities $\pm\omega_k$. In addition, $H$ is factorized as
(see Eq.(\ref{teorema g0}))
\begin{equation} \label{hamiltoniano gral imaginarios}
    H = \sum_{k=1}^{n} \gamma_k \omega_k L_k^+ L_k^- + g_0,
    \qquad g_0\in{\mathbb R}.
\end{equation}
Notice that the previous summation either involves terms for which ${\gamma_k} = 1$,
which are of the oscillator kind since they are positively defined, or terms with ${\gamma_k}
= - 1$ which are of the {\it anti-oscillator} type since they are negatively defined.
Thus, it is natural to define a {\it global algebraic structure} for our system (see, e.g.,
\cite{fhr07,fv09}), which is independent of the spectral details but has to do with the fact
that any quadratic Hamiltonian in the trap regime can be expressed in terms of several
independent oscillators, some of them indeed being anti-oscillators (compare
Eq.(\ref{hamiltoniano gral imaginarios})). This global structure is characterized mathematically
by identifying $n$ sets of number, annihilation and creation operators of the system,
$\{N_k,B_k,B_k^\dagger\}, \ k=1,\dots,n$, in the way:
\begin{eqnarray}\label{definicion B's}\nonumber
   B_k &=& L_k^-,\qquad  B_k^\dagger=L_k^+,\qquad
     \textnormal{for} \quad \gamma_k=1,\\
   B_k &=& L_k^+,\qquad  B_k^\dagger=L_k^-,\qquad
     \textnormal{for}\quad \gamma_k=-1, \\
   N_k &=& B_k^\dagger B_k, \quad  k=1,\dots,n,  \nonumber
\end{eqnarray}
so that the standard commutation relationships are satisfied,
\begin{eqnarray}\label{conmutador bb+}
  & [B_j,B_k^\dagger] & = \delta_{jk}, \qquad [B_j , B_k ]  = [B_j ^\dagger , B_k ^ \dagger ] = 0, \\
  \nonumber  & [N_k , B_k ] & =  - B_k, \quad [N_k , B_k^\dagger ] = B_k^\dagger,
 \qquad j,k=1,\dots,n.
\end{eqnarray}

Let us construct now a basis $\{\vert n_1,\dots,n_n\rangle, n_j=0,1,2,\dots,\
j=1,2,\dots,n\}$ of common eigenstates of $\{N_1,\dots,N_n\}$ (the Fock states)
\begin{equation}
N_j \vert n_1,\dots,n_n \rangle = n_j \vert n_1,\dots,n_n \rangle,
\quad j=1,\dots,n,
\end{equation}
starting from an {\it extremal state} $\vert 0,\dots,0 \rangle$, which
is annihilated simultaneously by $B_1, \dots, \ B_n$:
\begin{equation}
B_j \vert 0,\dots,0 \rangle = 0, \quad j = 1,\dots,n.
\end{equation}
If we assume that $\vert 0,\dots,0 \rangle$ is normalized, it turns out
that:
\begin{equation}\label{Fockstates}
\vert n_1,\dots,n_n \rangle =
\frac{B_1^\dagger{}^{n_1} \dots B_n^\dagger{}^{n_n}\vert 0,\dots,0 \rangle}{
\sqrt{n_1! \cdots n_n!}} .
\end{equation}
Moreover, $B_j$ and $B_j^\dagger, \ j = 1,\dots,n$, act onto $\vert
n_1,\dots,n_n \rangle$ in a standard way:
\begin{eqnarray}\nonumber
\hskip-1cm &  B_j \vert n_1,\dots,n_{j-1},n_j,n_{j+1},\dots, n_n \rangle \! =  \!
\sqrt{n_j} \, \vert n_1 ,\dots,n_{j-1},n_j - 1,n_{j+1},\dots,n_n \rangle,  \\
\hskip-1cm & B_j^\dagger \vert
n_1,\dots,n_{j-1},n_j,n_{j+1},\dots,n_n \rangle  \! = \!\! \sqrt{n_j \!\! + \! 1} \vert n_1,
\dots,n_{j-1},n_j \!\! + \! 1,n_{j+1},\dots,n_n \rangle .
\end{eqnarray}

Now, in terms of the operators $\{B_j, \, B_j^\dagger, \, j= 1,
\dots,n\}$ our Hamiltonian is expressed by
\begin{equation}  \label{H en B's}
H = \sum_{k=1}^n \gamma_k \omega_k B_k^\dagger B_k + g'_0.
\end{equation}
It is clear that the Fock states $\vert n_1,\dots,n_n \rangle$ are
eigenstates of $H$ with eigenvalues $E_{n_1,\dots,n_n} = \gamma_1\omega_1 n_1
+ \dots + \gamma_n \omega_n n_n + g_0' \equiv E(n_1,\dots,n_n)$. In particular,
the extremal state $\vert 0,\dots,0 \rangle$ has eigenvalue $E_{0,\dots,0} = g'_0$.
In case that ${\gamma_k} = 1$ for all $k$, then $H - g_0'$ will be a positively
defined operator, and the extremal state $\vert 0,\dots,0 \rangle$ will become the
ground state for our system, associated to the lowest eigenvalue
$E_{0,\dots,0} = g'_0$ of $H$. On the other hand, if there is at
least one index $j$ for which $\gamma_j = -1$, then $H - g_0'$ will not be
positively defined, since the corresponding $j$-th term is of
{\it inverted oscillator} type, and the state $\vert 0,\dots,0 \rangle$ will
not be a ground state for our system (however it keeps its extremal nature
since it is always annihilated by the $n$ operators $B_j, j=1,\dots,n$).

Following \cite{fhr07,fv09} it is straightforward to see that, besides the global
algebraic structure, there is an {\it intrinsic algebraic structure} for our system,
characterized by the existing relationship between the Hamiltonian $H$ and the
$n$ number operators $N_k$:
\begin{equation}
H = E(N_1,\dots,N_n) = \sum\limits_{k=1}^n \gamma_k \omega_k N_k + g'_0. \label{intrinsicE(N)}
\end{equation}
As in the examples discussed in \cite{fhr07,fv09}, it turns out that this intrinsic
algebraic structure is responsible for the specific spectrum of our Hamiltonian. On
the other hand, the global algebraic structure arises from the existence of the $n$
independent oscillator modes for $H$, each
one characterized by the standard generators $\{N_j,B_j,
B_j^\dagger\}, \ j = 1,\dots,n$. This global behavior allows us to identify
in a natural way the extremal state $\vert 0,\dots,0\rangle \equiv\vert{\mathbf 0}\rangle$,
which plays the role of a ground state although it does not necessarily
has a minimum energy eigenvalue. Moreover, the very existence of the extremal state
$\vert{\mathbf 0}\rangle$ is guaranteed by a theorem \cite{camporotante} ensuring that
if the operators $\{B_1,\dots,B_n\}$ obey the commutation relations given by
Eq.(\ref{conmutador bb+}), then the system of
partial differential equations
\begin{equation}\label{sistema de ecuaciones}
    \langle\vec{x}|B_j \vert 0,\dots,0\rangle =
    \langle\vec{x}|B_j \vert {\mathbf 0}\rangle = 0,
    \quad
    j=1,\dots,n,
\end{equation}
has the square integrable solution
\begin{equation} \label{solucion sistema de ec}
   \phi_{\mathbf 0}(\vec{x}) =  \langle \vec{x} \vert {\mathbf 0} \rangle
   = c e^{-\frac{1}{2}a_{ij}x_i x_j}=c
    e^{-\frac{1}{2}(\vec{x}^T\mathbf{a}\vec{x})},
\end{equation}
with $c$ being a normalization factor. In this wave function, $\mathbf{a}=(a_{ij})$
represents a symmetric matrix whose complex entries are found by solving the
system of equations
(\ref{sistema de ecuaciones}), leading to
\begin{equation} \label{alfa a igual a beta}
    \mathbf{a}\vec{\alpha}_j=\vec{\beta}_j, \quad
    j=1,\dots,n,
\end{equation}
where $\vec{\alpha}_j$ and $\vec{\beta}_j$ are obtained by
expressing $B_j$ and $B_j^\dagger$ as
\begin{equation} \label{desarrollo B en alfa y beta}
    B_j=i\vec{P}\cdot\vec{\alpha}_j + \vec{X}\cdot\vec{\beta}_j,
    \quad B_j^\dagger=-i\vec{\alpha}_j^\dagger \cdot \vec{P} + \vec{\beta}_j^\dagger \cdot
    \vec{X},\quad j=1,\dots,n.
\end{equation}
The wave functions for the other Fock states can be found from Eq.(\ref{Fockstates}).

\section{Coherent States }\label{C:Cap5}

Once our Hamiltonian has been expressed appropriately in terms of annihilation and creation
operators, we can develop a similar treatment as for the harmonic oscillator to build up the
corresponding coherent states. Here we are going to construct them either as simultaneous
eigenstates of the annihilation operators of the system or as the ones resulting from acting
the global displacement operator onto the extremal state.

\subsection{Annihilation Operator Coherent States (AOCS)}

In the first place let us look for the annihilation operator
coherent states (AOCS) as common eigenstates of the $B_j$'s:
\begin{equation}
B_j \vert
z_1,\dots,z_n \rangle = z_j \vert z_1,\dots,z_n \rangle, \quad z_j\in{\mathbb C} ,
\quad j=1,\dots,n. \label{cseB}
\end{equation}
Following a standard procedure, let us expand them in the basis $\{
\vert n_1,\dots,n_n \rangle \}$:
\begin{equation}
\vert z_1,\dots,z_n \rangle = \sum\limits_{n_1,\dots,n_n = 0}^\infty
c_{n_1,\dots,n_n} \vert n_1,\dots,n_n \rangle.
\end{equation}
By imposing now that Eq.(\ref{cseB}) is satisfied, the following recurrence
relationships are obtained,
\begin{equation}
c_{n_1,\dots,n_j,\dots, n_n} = \frac{z_j}{\sqrt{n_j}}c_{n_1,\dots,n_j-1,\dots, n_n},
\quad j=1,\dots,n ,
\end{equation}
which, when iterated, lead to
\begin{equation}
c_{n_1,\dots,n_j,\dots, n_n} = \frac{z_j^{n_j}}{\sqrt{n_j!}}
\, c_{n_1,\dots,0,\dots, n_n}, \quad j = 1,\dots,n.
\end{equation}
Hence, it turns out that
\begin{equation}
c_{n_1,\dots,n_n} =
\frac{z_1^{n_1}\dots z_n^{n_n}}{\sqrt{n_1!\cdots n_n!}} \, c_{0,\dots,0},
\end{equation}
where $c_{0,\dots,0}$ is to be found from the normalization condition.
Thus the normalized AOCS become
finally:
\begin{equation}
\vert z_1,\dots,z_n \rangle = \exp\left(-\frac12\sum_{j=1}^n
\vert z_j\vert^2 \right)
\sum_{n_1,\dots,n_n = 0}^\infty
\frac{z_1^{n_1}\dots z_n^{n_n} \vert n_1,\dots,n_n \rangle}{\sqrt{n_1! \cdots
n_n!}} , \label{cspta}
\end{equation}
up to a global phase factor.

\subsection{Displacement Operator Coherent States (DOCS)}\label{subsection
Desplazamiento}

The displacement operator for the $j$-th oscillator mode of the Hamiltonian reads
\begin{equation}
D_j(z_j)=\exp\left(z_jB_j^\dagger-z_j^*B_j\right).
\end{equation}
By using the BCH formula it turns out that
\begin{equation}
D_j(z_j)=\exp\left(-\frac{\vert z_j\vert^2}{2}\right)
\exp\left(z_j B_j^\dagger\right)\exp\left(-z_j^*B_j\right).
\end{equation}
Now, the global displacement operator is given by:
\begin{equation}
D(\mathbf{z}) \equiv D(z_1,\dots,z_n)= D_1(z_1)\cdots D_n(z_n),
\end{equation}
where $\mathbf{z}$ denotes the complex variables $z_1,\dots, z_n$
associated to the $n$ oscillator modes.

Let us obtain now the displacement operator coherent states (DOCS)
$|{\mathbf{z}}\rangle$ from applying $D(\mathbf{z})$ onto the extremal
state $\vert 0,\dots,0 \rangle \equiv\vert{\mathbf 0}\rangle$:
\begin{equation}\label{z de D en z=0}
|\mathbf{z}\rangle = D({\mathbf{z}})| {\mathbf 0} \rangle = \exp\left(-\frac12\sum_{j=1}^n
\vert z_j\vert^2 \right) \sum_{n_1,\dots,n_n = 0}^\infty
\frac{z_1^{n_1}\dots z_n^{n_n} \vert n_1,\dots,n_n \rangle}{\sqrt{n_1! \cdots
n_n!}} .
\end{equation}
Notice that the AOCS and the DOCS are the same (compare Eqs.(\ref{cspta}) and
(\ref{z de D en z=0})).

\subsection{Coherent state wave functions}

In order to find the wave functions of the coherent states previously derived, we
employ that $[z_jB_j^\dagger - z_j^* B_j,z_k B_k^\dagger-z_k^* B_k]=0 \ \forall
\ j,k$. Thus:
\begin{eqnarray}\nonumber
D(\mathbf{z}) &=& \exp(z_1 B_1^\dagger-z_1^*B_1)
    \cdots \exp(z_n B_n^\dagger-z_n^*B_n) \\
    & = & \exp[(z_1 B_1^\dagger + \cdots + z_n B_n^\dagger)
    -(z_1^*B_1 + \cdots + z_n^*B_n)].
\end{eqnarray}
Using now Eq.(\ref{desarrollo B en alfa y beta}) we can write
\begin{equation}\label{op D en Gama y Sigma}
    D(\mathbf{z}) =  e^{-i(\vec{\Gamma}\cdot \vec{P}-\vec\Sigma\cdot\vec{X})}
     = e^{-\frac{i}{2}\vec{\Gamma}\cdot\vec{\Sigma}}
    e^{i\vec{\Sigma}\cdot\vec{X}}
    e^{-i\vec{\Gamma}\cdot\vec{P}}
    =e^{\frac{i}{2}\vec{\Gamma}\cdot\vec{\Sigma}}
    e^{-i\vec{\Gamma}\cdot\vec{P}}
    e^{i\vec{\Sigma}\cdot\vec{X}},
\end{equation}
where we have employed once again the BCH formula and we have taken
\begin{equation}\label{vec Gama y Sigma}
    \vec{\Gamma} = 2 {\rm{Re}}[z_1^*
    \vec{\alpha}_1 + \dots + z_n^*\vec{\alpha}_n], \quad
    \vec{\Sigma} = -2{\rm{Im}}[z_1^*\vec{\beta}_1 + \dots + z_n^*\vec{\beta}_n].
\end{equation}
Now, it is straightforward to find the wave function for the coherent
state $|\mathbf{z}\rangle$,
\begin{equation}\nonumber
\phi_\mathbf{z}(\vec{x}) = \langle \vec{x}|\mathbf{z}\rangle = \langle \vec{x}|D(\mathbf{z})|\mathbf{0} \rangle
= e^{-\frac{i}{2}\vec{\Gamma}\cdot\vec{\Sigma}} e^{i\vec{\Sigma}\cdot\vec{x}}
\langle \vec{x}| e^{-i\vec{P}\cdot\vec{\Gamma}} |\mathbf{0}\rangle .
\end{equation}
Since the operator $\vec{P}$ is the coordinate displacement generator
\cite{Cohen1}, it turns out that
\begin{equation}\nonumber
    \langle \vec{x}|e^{-i\vec{P}\cdot\vec{\Gamma}}= \langle
    \vec{x}-\vec{\Gamma}|,
\end{equation}
so that
\begin{equation}\label{eewf}
    \phi_\mathbf{z}(\vec{x})=
    e^{-\frac{i}{2}\vec{\Gamma}\cdot\vec{\Sigma}}
    e^{i\vec{\Sigma}\cdot\vec{x}}
    \langle
    \vec{x}-\vec{\Gamma}| \mathbf{0} \rangle
    = e^{-\frac{i}{2}\vec{\Gamma}\cdot\vec{\Sigma}}
    e^{i\vec{\Sigma}\cdot\vec{x}}
    \phi_\mathbf{0} ( \vec{x}-\vec{\Gamma}).
\end{equation}
A further calculation, using Eq.(\ref{solucion sistema de ec}), leads finally to
\begin{equation}\label{form funcional edo coherente}
    \phi_\mathbf{z}(\vec{x})
    = e^{-\frac{1}{2}(\vec{\Gamma}^T\mathbf{a}+i \vec{\Sigma})\cdot\vec{\Gamma}}
    e^{(\vec{\Gamma}^T\mathbf{a}+i\vec{\Sigma})\cdot\vec{x}}
    \phi_\mathbf{0}(\vec{x}).
\end{equation}
Once again, it becomes evident that the extremal state is important in our treatment,
since its wave function determines the corresponding wave function for any other CS.
Moreover, as it can be seen from Eq.(\ref{eewf}), the position probability density
for the CS $|\mathbf{z}\rangle$ becomes just a displaced version of the corresponding
one for the extremal state $|\mathbf{0}\rangle$.

\section{Mathematical and physical properties}

Let us derive next the completeness relationship for the previously derived coherent
states. Notice that, from the point of view of the analysis of states in the Hilbert
space of the system, this is the most important property which our CS would have
\cite{kl63a,kl63b,libroCompliacion,NuevoDelProfe}. This is the reason why several
authors use it as the fourth coherent state definition, considering it as the fundamental
one which will survive in time (see e.g. \cite{NuevoDelProfe}). We are going to calculate
as well some important physical quantities in these states.

\subsection{Completeness relationship}

A straightforward calculation leads to:
\begin{eqnarray}
   &&  \hskip-2.3cm \left(\frac1{\pi}\right)^n\int\cdots\int |\mathbf{z} \rangle\langle \mathbf{z}
  | d^2 z_1\dots d^2 z_n
    \nonumber \\
   && \hskip-1.2cm  = \!\! \!\! \sum_{m_1, n_1,\dots, m_n, n_n = 0}^\infty \frac{
    |m_1,\dots, m_n\rangle\langle n_1,\dots, n_n |}{\sqrt{m_1! n_1!\cdots m_n! n_n!}}
    \prod_{j=1}^{n} \left(\frac1{\pi} \int z_j^{m_j}{z_j^*}^{n_j}
     e^{- \vert z_j\vert^2}d^2z_j \right) = {\bf 1},
\end{eqnarray}
with ${\bf 1}$ being the identity operator. Thus, the coherent states
$\{| \mathbf{z} \rangle\}$ form a complete set in the state space
of the system (indeed they constitute an overcomplete set
\cite{sobrecompletez,AcercaDeLaCompletez}). This implies that any state
can be expressed in terms of our coherent states,
in particular, an arbitrary coherent state,
\begin{equation}
| \mathbf{z'} \rangle = \left(\frac1{\pi}\right)^n\int\cdots\int |\mathbf{z}
\rangle\langle \mathbf{z} | \mathbf{z'} \rangle d^2 z_1\dots d^2 z_n,
\end{equation}
where the reproducing kernel $\langle \mathbf{z} | \mathbf{z'} \rangle$ is
given by
\begin{equation}
\langle \mathbf{z} | \mathbf{z'} \rangle =
\exp\left[ - \frac12 \sum_{j=1}^n \left( |z_j|^2 - 2z_j^*z_j'  + |z'_j|^2\right)
\right].
\end{equation}
This means that, in general, our coherent states are not orthogonal to each other. Indeed,
notice that inside our infinite set of coherent states only the extremal state of the
system, $|\mathbf{0}\rangle\equiv|\mathbf{z}=\mathbf{0}\rangle=|0,\dots,0\rangle$, is also
an eigenstate of the Hamiltonian.

\subsection{Mean values of some physical quantities in a CS}\label{S:cantidades fisicas}

Now we can calculate easily the mean values $\langle X_j\rangle_\mathbf{z}$
$\equiv$ $\langle \mathbf{z}|X_j|\mathbf{z}\rangle$, $\langle
P_j\rangle_\mathbf{z}\equiv \langle
\mathbf{z}|P_j|\mathbf{z}\rangle$, $j=1,\dots,n$, in a given coherent state
$|\mathbf{z}\rangle=|z_1,\dots,z_n\rangle$, as well as its mean
square deviation in terms of the corresponding results for the
extremal state $| \mathbf{0} \rangle$. To do that, let us analyze first
how the operators $X_j, \ X_j^2, \ P_j, \ P_j^2$ are transformed under
$D(\mathbf{z})$. By using Eqs.(\ref{op D en Gama y Sigma},\ref{vec Gama y Sigma})
it is straightforward to show that:
\begin{equation}
\hskip-1cm D^\dagger(\mathbf{z}) X_j^n D(\mathbf{z}) =
(X_j + \Gamma_j)^n, \quad D^\dagger(\mathbf{z}) P_j^n
D(\mathbf{z}) =  (P_j + \Sigma_j)^n, \quad n = 1,2,\dots
\end{equation}
where we have used that, for an operator $A$ which commutes with $[A,B]$,
it turns out that
\begin{equation}\label{operador entre exponenciales}
\hskip-0.5cm e^{A}Be^{-A}=B + [A,B] \quad \Rightarrow \quad
e^{A}B^n e^{-A}=(B + [A,B])^n, \quad n=1,2,\dots.
\end{equation}
Thus, a straightforward calculation leads to
\begin{equation}\label{valor esperado X_j}
    \langle X_j\rangle_\mathbf{z} \equiv \langle \mathbf{z}|X_j|\mathbf{z}\rangle
    =  \langle \mathbf{0} |D^\dagger(\mathbf{z}) X_j D(\mathbf{z})| \mathbf{0} \rangle
    = \langle X_j \rangle_\mathbf{0} + \Gamma_j.
\end{equation}
On the other hand,
\begin{equation}
\label{valor esperado X^2_j}
    \langle X_j^2\rangle_\mathbf{z} =
    \langle X_j^2\rangle_\mathbf{0} + 2\Gamma_j\langle
    X_j\rangle_\mathbf{0} + {\Gamma_j}^2.
\end{equation}
Hence,
\begin{equation}\label{Delta X_z = Delta X_0}
    (\Delta X_j)^2_\mathbf{z}=\langle X_j^2 \rangle_\mathbf{z} -\langle X_j
    \rangle_\mathbf{z}^2
    =\langle X_j^2 \rangle_\mathbf{0} -\langle X_j\rangle_\mathbf{0}^2
    =(\Delta X_j)_\mathbf{0}^2.
\end{equation}

Working in a similar way for $P_j$, it is obtained
\begin{equation}\label{valor esperado P_j}
  \langle P_j \rangle_\mathbf{z} =  \langle P_j\rangle_\mathbf{0} + \Sigma_j, \qquad
  \langle P_j^2\rangle_\mathbf{z}
   = \langle P_j^2\rangle_\mathbf{0} + 2\Sigma_j\langle
    P_j\rangle_\mathbf{0} + {\Sigma_j}^2.
\end{equation}
Then we have as well that
\begin{equation} \label{Delta P_z = Delta P_0}
    (\Delta P_j)_\mathbf{z}^2=(\Delta P_j)_\mathbf{0}^2,
\end{equation}
i.e., the mean square deviations of $X_j$ and $P_j$ in the CS
$|\mathbf{z}\rangle$ are independent of $\mathbf{z}$.

In order to end up this calculation, the mean values $\langle
X_j\rangle_\mathbf{0}$, $\langle P_j\rangle_\mathbf{0}$,
$\langle X_j^2\rangle_\mathbf{0}$, and $\langle P_j^2\rangle_\mathbf{0}$
for $j=1,\dots,n$ are required. Let us describe now the procedure to find
these $4n$ quantities. The first $2n$, $\langle X_j\rangle_\mathbf{0}$,
$\langle P_j\rangle_\mathbf{0}$, can be easily found by recalling the definitions
of $B_k$, $B_k^\dagger$ (see section 2.1) and using the
fact that their mean values in the extremal state $| \mathbf{0} \rangle$ always
vanish for $k = 1,\dots,n$:
\begin{eqnarray}\label{linear system}
&&   \langle B_k \rangle_\mathbf{0} = \langle B_k^\dagger \rangle_\mathbf{0} = 0, 
\quad k=1,\dots,n.
\end{eqnarray}
This is equivalent to the following linear system of $2n$ homogeneous equations 
\begin{eqnarray}
&& f_k^- \langle \eta \rangle_\mathbf{0} = f_k^+ \langle \eta \rangle_\mathbf{0} = 0, 
\quad k=1,\dots,n.
\end{eqnarray}
Since the left eigenvectors $f_k^\pm, k=1,\dots,n$, are linearly independent, the only 
solution for the $2n$ unknowns $\langle \eta \rangle_\mathbf{0}$ is the trivial one,
i.e., $\langle X_j\rangle_\mathbf{0} = \langle P_j\rangle_\mathbf{0}=0$. It is worth to point
out that this result simplifies 
Eqs.(\ref{valor esperado X_j},\ref{valor esperado X^2_j},\ref{valor esperado P_j}).

On the other hand, the mean values of the quadratic operators
$X_i^2$, $P_i^2, \ i = 1,\dots,n$, in the extremal state $|\mathbf{0}\rangle$
can be obtained from evaluating the corresponding quantities for the several
non-equivalent products of pairs of annihilation $B_j$ and creation $B_k^\dagger$
operators. It is important to mention that these products should have the appropriate
order to use the fact that $B_j$ annihilates $| \mathbf{0} \rangle$ and $B_k^\dagger$
annihilates $\langle \mathbf{0}|$ (if the product involves one $B_j$ it should be
placed to the right while if it involves one $B_k^\dagger$ it should be placed to the
left). In general, we will get $n(2n+1)$ non-equivalent products of pairs of operators
$B_i$, $B_j^\dagger$, $i,j=1,\dots,n$: $n(n+1)/2$ products of kind $B_i B_j, j=1,\dots,n,
\ i\leq j$; $n(n+1)/2$ products of kind $B_i^\dagger B_j^\dagger, j=1,\dots,n, \ i\leq j$;
$n^2$ products of kind $B_i^\dagger B_j, \ i,j=1,\dots,n$.
The mean values in the extremal state $|\mathbf{0}\rangle$ will lead to an inhomogeneous
systems of $n(2n+1)$ equations, with the same number of unknowns.
When solving this system we will get $\langle X_j^2\rangle_\mathbf{0}$,
$\langle P_j^2\rangle_\mathbf{0}$, $j=1,\dots,n$, and the mean
value of any other product of two canonical operators $X_i$, $P_j$.

It is customary nowadays to group the mean values of the quadratic products of the 
operators $X_i, \ P_j$ in the coherent state $|\mathbf{z}\rangle$ in a $2n \times 2n$
real symmetric matrix $\sigma(\mathbf{z})$, called covariance matrix, whose elements
are given by (remember that $\eta = (X_1, \dots, X_n, P_1, \dots, P_n)^T$):
\begin{eqnarray}\label{mecmz}
& \sigma_{ij}(\mathbf{z}) = \frac12\langle \eta_i \eta_j + \eta_j \eta_i
\rangle_{\mathbf{z}} - \langle \eta_i \rangle_{\mathbf{z}} \langle \eta_j \rangle_{\mathbf{z}},
\quad i,j=1,\dots 2n.
\end{eqnarray}
A straightforward calculation leads to
\begin{eqnarray}\label{mecm0}
& \sigma_{ij}(\mathbf{z}) = \sigma_{ij}(\mathbf{0}) = \frac12\langle \eta_i \eta_j 
+ \eta_j \eta_i \rangle_{\mathbf{0}} \equiv \sigma_{ij} ,
\end{eqnarray}
where we have used that $\langle \eta_i \rangle_{\mathbf{0}} = 0, \ i=1,\dots,2n$. The conclusion
is that the covariance matrix in our coherent state $|\mathbf{z}\rangle$ is once again independent
from $\mathbf{z}$ and depends just of the extremal state $|\mathbf{0}\rangle$. 
Notice that the number of independent matrix elements $\sigma_{ij}$ ($n(2n+1)$) coincides
with the number of unknowns which are determined from the set of $n(2n+1)$ independent equations 
associated to the mean values of the quadratic products of $B_j^\dagger, B_k$ in the extremal state
$|\mathbf{0}\rangle$.

Once the covariance matrix is determined, the generalized uncertainty relation 
can be evaluated \cite{ha96,ai07,gl09b}
\begin{eqnarray}\label{rsur}
& \sigma_{i \, i} \, \sigma_{n+i \, n+i} - \sigma_{i \, n+i}
\geq \frac14, \quad i=1,\dots,n,
\end{eqnarray}
which coincides with the Robertson-Schr\"odinger uncertainty relation (see e.g. \cite{ha96}).

Let us end up this section by calculating the mean value of the Hamiltonian
in a given coherent state $|\mathbf{z}\rangle$. Equation (\ref{H en B's})
leads to
\begin{equation}
    \langle H \rangle_\mathbf{z} = \langle \mathbf{z}|H|\mathbf{z}
    \rangle=\sum_{k=1}^n \gamma_k \omega_k |z_k|^2 + g'_0.
\end{equation}
In order to get $\langle H^2\rangle_{\mathbf{z}}$, let us notice that
\begin{equation} \nonumber
 \hskip-1.0cm   H^2 = \sum_{j,k=1}^n\gamma_j\gamma_k\omega_j\omega_kB_j^\dagger B_k^\dagger
    B_j   B_k+\sum_{k=1}^n \omega_k^2 B_k^\dagger B_k
    + 2g'_0\sum_{k=1}^n \gamma_k \omega_k B_k^\dagger B_k +{g'_0}^2.
\end{equation}
Thus we get
\begin{equation} \nonumber
 \hskip-1.0cm   \langle H^2\rangle_\mathbf{z}=\sum_{j,k=1}^n\gamma_j\gamma_k\omega_j\omega_k
    |z_j|^2|z_k|^2
    +\sum_{k=1}^n \omega_k^2 |z_k|^2
    + 2g'_0\sum_{k=1}^n \gamma_k \omega_k |z_k|^2 +{g'_0}^2.
\end{equation}
Hence,
\begin{equation} \label{Delta H}
    (\Delta H)^2_\mathbf{z} = \sum_{k=1}^n\omega_k^2 |z_k|^2.
\end{equation}

Notice that, for one-dimensional systems ($n=1$), this expression reduces to the standard
one for the harmonic oscillator (see e.g. \cite{Cohen1}).

\subsection{Time evolution of the CS}\label{time-evolution}

Suppose that at $t=0$ our system is in a coherent state $|\mathbf{z}\rangle$.
Thus, at a later time $t>0$ the evolved state is found by acting on $|\mathbf{z}\rangle$
with the evolution operator of the system $U(t) = \exp(-iHt)$. By making use of
Eq.(\ref{intrinsicE(N)}) it turns out that
$$
U(t) = e^{-ig_0' t} \prod_{k=1}^{n} e^{-i\gamma_k \omega_k N_k} .
$$
Hence,
\begin{equation}\label{coherenteevolucionado}
U(t) |\mathbf{z}\rangle = e^{-ig_0' t} |z_1(t),\dots, z_n(t)  \rangle = e^{-ig_0' t}
|\mathbf{z}(t)\rangle,
\end{equation}
where $z_j(t) = e^{-i\gamma_j \omega_j t} z_j = \vert z_j\vert e^{i(\theta_j-\gamma_j \omega_j t)}$. 
Equation (\ref{coherenteevolucionado}) implies that a
coherent state $|\mathbf{z}\rangle$ evolves in time into a new coherent state $|\mathbf{z}(t)\rangle =
|z_1(t),\dots, z_n(t)\rangle$, where the $j$-th degree of freedom $z_j(t)$ just rotates at its characteristic
frequency $\omega_j$ (clockwise if $\gamma_j = 1$ and counterclockwise if $\gamma_j = -1$).

\subsection{Gazeau-Klauder coherent states}\label{Gazeau-Klauder}

At this point, it would be interesting to check if our CS belong to the class introduced
recently by Gazeau and Klauder \cite{gk99}. Using their notation, for a system with a Hamiltonian
${\cal H}$ such that the ground state energy is zero, the Gazeau-Klauder CS $\{\vert J,\theta
\rangle, \ J\geq 0, \ -\infty < \theta < \infty\}$ obey the 
following properties:

\noindent (a) Continuity: $(J',\theta') \rightarrow (J,\theta) \Rightarrow \vert J',\theta'
\rangle \rightarrow \vert J,\theta\rangle$.

\noindent (b) Resolution of unity: ${\bf 1} = \int \vert J,\theta\rangle \langle J,\theta \vert
d\mu(J,\theta)$.

\noindent (c) Temporal stability: $e^{-i {\cal H}t}\vert J,\theta\rangle = \vert J,\theta + \omega t\rangle,
\ \omega = {\rm constant}$.

\noindent (d) Action identity: $\langle J,\theta \vert {\cal H} \vert J,\theta\rangle = \omega J$.

Concerning the first property, it is straightforward to check that our coherent states given in 
Eq.(\ref{z de D en z=0}) are such that $|\mathbf{z'}\rangle \rightarrow
|\mathbf{z}\rangle$ as $\mathbf{z'}\rightarrow\mathbf{z}$, i.e., they are continuous in $\mathbf{z}$.
As for the second and third properties, both were explicitly proven in sections 4.1 and 4.3 respectively.
It remains just to analyze if it is valid the action identity given in (d). Let us notice first of 
all that it is valid for each partial Hamiltonian $H_k = \gamma_k\omega_k N_k$ of our system,
$$
\langle \mathbf{z} \vert H_k \vert \mathbf{z}\rangle = \gamma_k\omega_k \vert z_k \vert^2,
$$
which is time-independent. Therefore, property (d) becomes valid for each degree of
freedom separately and thus it is valid for our global system with the natural identification
$J_k = \vert z_k \vert^2$, $\theta_k = {\rm arg}(z_k)$ so that
$$
\langle \mathbf{z} \vert (H - g_0') \vert \mathbf{z}\rangle = \sum_{k=1}^{n}\gamma_k \omega_k J_k.
$$
We conclude that our CS of Eq.(\ref{z de D en z=0}) become as well an $n$-dimensional generalization of
the Gazeau-Klauder CS if we express each complex component $z_k$ of $\mathbf{z}$ in its polar form 
(the polar coordinates essentially coincide with the canonical action-angle variables for the 
corresponding classical system).

\section{Asymmetric Penning trap coherent states}{}

Let us apply now the previous technique to the asymmetric Penning trap.
Such an arrangement can be used to control some quantum mechanical phenomena
\cite{fm94} as well as to perform high-precision measurements
of fundamental properties of particles. Moreover, it is a quite natural
system to analyze the decoherence
taking place due to the unavoidable interaction of the system with its environment
\cite{bg86,jl09}. Since the asymmetric Penning trap becomes the ideal one when the
asymmetry parameter vanishes \cite{fn91,fe92,fb93}, it will be straightforward to
compare these results with those recently obtained for the ideal Penning trap
\cite{fv09} (see also \cite{ma97,chlm98,gl09b}).

The Hamiltonian of a charged particle with mass $m$ and charge $q$
in an asymmetric Penning trap reads
\begin{equation}
H = \frac{{\vec P}^2}{2m} + \frac{\omega_c}{2} (XP_y-YP_x) + \frac{m}{2}
(\omega_x^2  X^2 +\omega_y^2 Y^2+\omega_z^2 Z^2),
\end{equation}
$\omega_c=qB/m$ and $\omega_z$ being the cyclotron and axial frequencies respectively,
and the frequencies $\omega_x, \omega_y$ are given by
\begin{equation}
\omega_x^2 = \frac{\omega_c^2}{4} - \frac{\omega_z^2}{2}(1+\varepsilon), \quad
\omega_y^2 = \frac{\omega_c^2}{4} - \frac{\omega_z^2}{2}(1-\varepsilon),
\end{equation}
where $\vert\varepsilon \vert < 1$ is the real asymmetry parameter and we are denoting
${\vec P} = (P_x,P_y,P_z)^T, {\vec X} = (X,Y,Z)^T$. Without loosing
generality \cite{fv09}, from now on we will assume that $m = 1$.

As it was seen at section 2, the main role in our treatment is played by the matrix
$\mathbf{\Lambda}$ such that $\left[iH,\eta\right]=\mathbf{\Lambda} \eta$. We choose here
$\eta=\left(X,Y,P_x,P_y,Z,P_z\right)^T$ so that
\begin{equation}
\mathbf{\Lambda} =  \left(
\matrix{
0 & -\omega_c /2 & 1 & 0 & 0 & 0 \cr
\omega_c/2 & 0 & 0 & 1 & 0 & 0 \cr
-\omega_x^2 & 0 & 0 & -\omega_c/2 & 0 & 0 \cr
0 & -\omega_y ^2 & \omega_c/2 & 0 & 0 & 0 \cr
0 & 0 & 0 & 0 & 0 & 1 \cr
0 & 0 & 0 & 0 & -\omega_z^2 & 0
}
\right) .
\end{equation}

The eigenvalues $(\lambda)$ of $\mathbf{\Lambda}$ are
\begin{eqnarray}
\lambda_1 & = & \frac{i\omega_c}{2} \sqrt{2 - \delta + R}= i\omega_1, \qquad
\lambda_2 = \frac{i\omega_c}{2} \sqrt{2-\delta - R} = i\omega_2,  \nonumber \\
\lambda_3 & = & i\omega_z= i\omega_3, \qquad  R = \sqrt{4(1-\delta)+\delta^2\varepsilon ^2}, \qquad 0 < \delta= \frac{2\omega_z^2}{\omega_c^2} < 1,
\end{eqnarray}
and their corresponding complex conjugate. The right $(u)$ and left $(f)$ eigenvectors of $\mathbf{\Lambda}$ become
\begin{eqnarray}
&& \hskip-1.5cm u_1^+ = s_1\left( {4\over \omega_c(\delta\varepsilon+R)}, -{i \over \omega_1} {2+\delta\varepsilon+R \over \delta\varepsilon+R}, {i\omega_c \over 2\omega_1} {2(1-\delta)-\delta\varepsilon+R\over \delta\varepsilon + R},1,0,0 \right)^T, \nonumber \\
&& \hskip-1.5cm u_2^+ = s_2\left( {4\over \omega_c(\delta\varepsilon-R)}, -{i \over \omega_2} {2+\delta\varepsilon-R \over \delta\varepsilon-R}, {i\omega_c \over 2\omega_2} {2(1-\delta)-\delta\varepsilon-R\over \delta\varepsilon - R},1,0,0 \right)^T, \nonumber \\
&& \hskip-1.5cm u_3^+ = s_3\left( 0,0,0,0,-{i\over\omega_z},1 \right)^T, \nonumber \\
&& \hskip-1.5cm f_1^+ = t_1\left( {\omega_c\over 4}(R-\delta\varepsilon),  {i\omega_c^2\over 8\omega_1} [2(1-\delta)+\delta\varepsilon+R] , -{i \omega_c \over 4\omega_1}(2-\delta\varepsilon+R),1,0,0 \right),\nonumber \\
&& \hskip-1.5cm f_2^+ = t_2\left( {\omega_c\over 4}(-R-\delta\varepsilon),  {i\omega_c^2\over 8\omega_2} [2(1-\delta)+\delta\varepsilon-R] , -{i \omega_c \over 4\omega_2}(2-\delta\varepsilon-R),1,0,0 \right),\nonumber\\
&& \hskip-1.5cm f_3^+ = t_3\left( 0,0,0,0,i\omega_z,1 \right), \label{rleapt}
\end{eqnarray}
where $s_j, t_j \in {\mathbb C}, j=1,2,3$. The requirement that the right and left eigenvectors be dual to each other implies
\begin{equation}
s_1={1\over 4t_1}\left(1+{\delta\varepsilon \over R}\right), \quad
s_2={1\over 4t_2}\left(1-{\delta\varepsilon \over R}\right), \quad
s_3={1\over 2t_3}.
\end{equation}
On the other hand, up to some phase factors, the condition imposed by
Eqs.(\ref{segundacondicion},\ref{modulogamma}) leads to,
\begin{equation}
t_1 = \frac{1}{\sqrt{2i(f_{1a}f_{1c}-f_{1b})}}, \quad t_2 = \frac{1}{\sqrt{2i(f_{2b}-
f_{2a}f_{2c})}}, \quad t_3 = \frac{1}{\sqrt{2\omega_3}},
\end{equation}
where we are denoting $f_1^+ = t_1(f_{1a}, f_{1b}, f_{1c}, 1,0,0),
f_2^+ = t_2(f_{2a}, f_{2b}, f_{2c}, 1,0,0)$ in order to simplify the
notation (compare Eq.(\ref{rleapt})). Moreover, the crucial signs for us to conclude that
our asymmetric Penning trap Hamiltonian is not positively defined become:
\begin{equation}
\gamma_1 = 1, \quad \gamma_2 = - 1, \quad \gamma_3 = 1.
\end{equation}
Thus, our annihilation operators take the form (see Eq.(\ref{definicion B's})):
\begin{eqnarray}
B_1=L_1^-&=&t_1\left(f_{1a}X-f_{1b}Y-f_{1c}P_x+P_y\right),\nonumber \\
B_2=L_2^+&=&t_2\left(f_{2a}X+f_{2b}Y+f_{2c}P_x+P_y\right), \\
B_3=L_3^-&=&t_3\left(-i \omega_3 Z+P_z \right). \nonumber
\end{eqnarray}
From these operators and their hermitian conjugates, it is straightforward to
identify the $\alpha_j$ and $\beta_j, \ j=1,2,3$ which allow us to find the matrix
${\mathbf{a}}$ such that $\mathbf{a}\alpha_j = \beta_j$. Its matrix elements $a_{ij}$
become now:
\begin{eqnarray}\label{aij}
a_{11}& = &-i {f_{1a}-f_{2a} \over f_{1c}+f_{2c} }, \quad
a_{12}=a_{21} = i {f_{1b}+f_{2b} \over f_{1c}+f_{2c} }, \quad
a_{22}=i {f_{1c}f_{2b}-f_{2c}f_{1b} \over f_{1c}+f_{2c} }, \nonumber \\
a_{33}&=& \omega_3, \qquad \qquad a_{13}=a_{31}=a_{23}=a_{32} = 0 .
\end{eqnarray}
It can be shown that $a_{11},a_{22},a_{33} \in {\mathbb R}^+$ while
$a_{12}$ is purely imaginary. Thus the extremal state wave function of
Eq.(\ref{solucion sistema de ec}) acquires the form:
\begin{equation}
\phi_\mathbf{0}(\vec{x}) = c \exp \left( - {1\over 2}a_{11}x^2 - {1\over 2}a_{22}y^2 -
a_{12}xy\right) \exp \left(- {1\over 2}a_{33}z^2\right).
\end{equation}
The associated eigenvalue becomes $E_{0,0,0} = (\omega_1 - \omega_2 + \omega_3)/2$.

Concerning the coherent states $\vert z_1,z_2,z_3 \rangle$, the general treatment developed
in section 3 is straightforwardly applicable, and their explicit expressions are given by
Eq.(\ref{z de D en z=0}) with $n=3$. Their corresponding wave functions are given by
\begin{equation}
\phi_z(\vec x) = \langle \vec x|\mathbf{z}\rangle= e^{-i\vec\Gamma \cdot
\vec\Sigma/2}e^{i\vec\Sigma \cdot \vec x}\phi_0(x-\Gamma_1,y-\Gamma_2,z-\Gamma_3),
\nonumber
\end{equation}
where
\begin{equation}
\hskip-1.45cm \vec{\Gamma}= 2 \left(
\matrix{
i t_1 f_{1c} {\rm Re}\left[z_1 \right]-i t_2 f_{2c}
{\rm Re}\left[z_2 \right]\cr 
- t_1  {\rm Im}\left[z_1 \right] -t_2
{\rm Im}\left[z_2 \right] \cr
-t_3  {\rm Im}\left[z_3 \right]
}\right),
\ \
\vec{\Sigma}=2\left(
\matrix{ 
t_1 f_{1a} {\rm Im}\left[z_1 \right] + t_2 f_{2a} {\rm Im}\left[z_2 \right] \cr  
- i t_1 f_{1b} {\rm Re}\left[z_1 \right]+i t_2 f_{2b}
{\rm Re}\left[z_2 \right] \cr 
t_3 \omega_3 {\rm Re}\left[z_3 \right]
}\right).
\end{equation}

The mean values $\langle X_j \rangle_\mathbf{z}$, $\langle P_j \rangle_\mathbf{z}$, immediately follow from 
Eqs.(\ref{valor esperado X_j}, \ref{valor esperado P_j}) with $\langle X_j \rangle_\mathbf{0} = 
\langle P_j \rangle_\mathbf{0} = 0$, i.e., 
\begin{eqnarray}
& \langle X_j \rangle_\mathbf{z} = \Gamma_j, \quad \langle P_j \rangle_\mathbf{z} = \Sigma_j, \quad j=1,2,3.
\end{eqnarray}
As for the mean values of the quadratic operators in the extremal state, we have solved the system of equations 
arising from the null mean values of the products of pairs of annihilation $B_j$ and creation  $B_k^\dagger$ 
operators. We get
\begin{eqnarray}\label{quaddiag}
\langle X^2 \rangle_\mathbf{0}&=& {1 \over 2 a_{11}}, \qquad \langle P_x^2 \rangle_\mathbf{0}= {1\over 2} \left( a_{11}-{a_{12}^2\over a_{22}}\right), \nonumber \\
\langle Y^2 \rangle_\mathbf{0}&=& {1 \over 2 a_{22}}, \qquad \langle P_y^2 \rangle_\mathbf{0}= {1\over 2} \left( a_{22}-{a_{12}^2\over a_{11}}\right), \\
\langle Z^2 \rangle_\mathbf{0}&=& {1 \over 2 a_{33}}, \qquad \langle P_z^2 \rangle_\mathbf{0}= {1\over 2} a_{33},
\nonumber
\end{eqnarray}
and the crossed products
\begin{eqnarray}\label{quadoutdiag}
\langle XP_x \rangle_\mathbf{0} = {i \over 2}, \qquad & \langle XP_y \rangle_\mathbf{0} = {i \over 2} {a_{12} \over a_{11}}, \qquad & \langle XP_z \rangle_\mathbf{0} = 0, \nonumber \\
\langle YP_x \rangle_\mathbf{0} = {i \over 2} {a_{12} \over a_{22}} , \qquad & \langle YP_y \rangle_\mathbf{0} = {i \over 2}, \qquad & \langle YP_z \rangle_\mathbf{0} = 0,  \\
\langle ZP_x \rangle_\mathbf{0} = 0, \qquad & \langle ZP_y \rangle_\mathbf{0} = 0, \qquad & \langle ZP_z \rangle_\mathbf{0} = {i \over 2}. \nonumber
\end{eqnarray}
Therefore, using Eqs.(\ref{Delta X_z = Delta X_0}, \ref{Delta P_z = Delta P_0}) we get the Heisenberg uncertainty relationships
\begin{eqnarray}\label{Heisenberg}
(\Delta X)^2_\mathbf{z} (\Delta P_x)^2_\mathbf{z} &=& (\Delta Y)^2_\mathbf{z} (\Delta P_y)^2_\mathbf{z}={1\over 4}\left(1+{\vert a_{12}\vert^2 \over a_{11}a_{22} }\right) \geq {1 \over 4},  \\
(\Delta Z)^2_\mathbf{z} (\Delta P_z)^2_\mathbf{z}&=& {1 \over 4}, \nonumber
\end{eqnarray}
while Eq.(\ref{Delta H}) with $n=3$ gives the mean square deviation for the Hamiltonian.

Once we have calculated the mean values of the quadratic products given in Eqs.(\ref{quaddiag},\ref{quadoutdiag}), 
it is straightforward to evaluate the covariance matrix elements of Eq.(\ref{mecm0}). With the ordering $\eta = 
(X,Y,P_x,P_y,Z,P_z)^T$, it is obtained:
\begin{eqnarray}
\sigma = \left(
\matrix{
(\Delta X)^2_\mathbf{0} & 0 & 0 & \frac{ia_{12}}{2 a_{11}} & 0 & 0 \cr
0 & (\Delta Y)^2_\mathbf{0} & \frac{ia_{12}}{2 a_{22}} & 0 & 0 & 0 \cr
0 & \frac{ia_{12}}{2 a_{22}} & (\Delta P_x)^2_\mathbf{0} & 0 & 0 & 0 \cr
\frac{ia_{12}}{2 a_{11}} & 0 & 0 & (\Delta P_y)^2_\mathbf{0} & 0 & 0 \cr
0 & 0 & 0 & 0 & (\Delta Z)^2_\mathbf{0} & 0 \cr
0 & 0 & 0 & 0 & 0 & (\Delta P_z)^2_\mathbf{0} 
} \right).
\end{eqnarray}
Notice that this covariance matrix is non-diagonal. However, since $\sigma_{13} = \sigma_{24}
= \sigma_{56} = 0$, it turns out that the generalized uncertainty relations of Eq.(\ref{rsur}) reduce to the Heisenberg 
uncertainty relations given in Eq.(\ref{Heisenberg}).

A plot of $(\Delta X)_\mathbf{z} (\Delta P_x)_\mathbf{z}$ as a function of the parameters
$\varepsilon$ and $\delta$ is given in Fig.1. As it can be seen from
Eqs.(\ref{aij},\ref{Heisenberg}) and from Fig.1, the coherent states minimize the Heisenberg
uncertainty relationship for $\varepsilon = 0$, which coincides with the results recently
obtained for the ideal Penning trap \cite{fv09}. However, for $\varepsilon \neq 0$ it turns
out that $(\Delta X)_\mathbf{z} (\Delta P_x)_\mathbf{z}>1/2$.
Notice that the same plot will appear for
the uncertainty product $(\Delta Y)_\mathbf{z} (\Delta P_y)_\mathbf{z}$.

\begin{figure}[ht]
\centering \includegraphics[width=12cm]{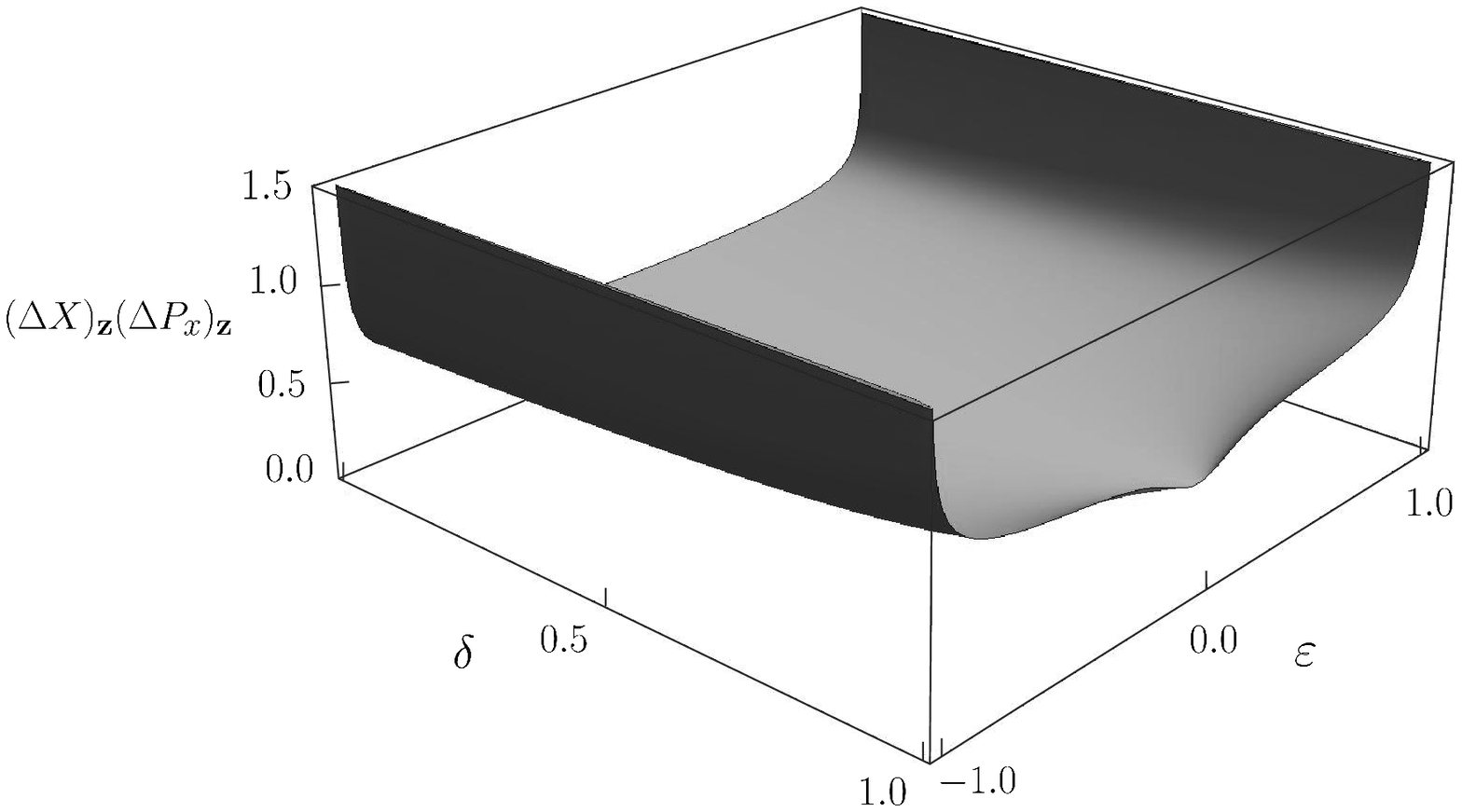}
\caption{\footnotesize Heisenberg uncertainty relationship
$(\Delta X)_\mathbf{z} (\Delta P_x)_\mathbf{z}$ for the asymmetric Penning trap
coherent states as function of the real dimensionless parameters $|\varepsilon |<1,
\ 0<\delta<1$.}
\end{figure}

\section{Concluding remarks}\label{C:conclusiones}

In this work we have proposed a systematic technique to find the CS for systems
governed by quadratic Hamiltonians in the trap regime. To do this, we introduced
a prescription to identify in a simple way the appropriate ladder operators which
play the same role as the annihilation and creation operators for the $1$-dimensional
harmonic oscillator. These operators allowed us to generate the
eigenvectors and eigenvalues for the Hamiltonian departing from the extremal
state, the analogue of the ground state although it is not necessarily
an eigenstate associated to the lowest possible eigenvalue. The explicit
expression for the extremal state wave function was as well explicitly calculated.

For systems governed by this kind of Hamiltonians the two algebraic CS definitions
(either as simultaneous eigenstates of the annihilation operators or as
resulting from the action of the displacement operator onto the extremal state)
lead to the same set of states. The explicit expression for the corresponding
wave functions has been also derived.

We have calculated explicitly the mean values of the position and momentum operators
in an arbitrary coherent state. Moreover, we have provided as well a prescription to obtain algebraically,
by solving a linear systems of equations, the mean values of the quadratic products
of these operators in the CS.

Through this method we have found the asymmetric Penning trap coherent states and we have explored
some of their physical properties. In particular, it is worth to point out that, in general, they do
not minimize the Heisenberg uncertainty relationship. The differences from the minimum are induced
by the deviations of the axial symmetry which the ideal Penning trap has (measured by the
asymmetry parameter $\varepsilon$).

\section*{Acknowledgments}{}

The authors acknowledge the support of Conacyt.

\end{document}